\documentclass[twocolumn,prd,superscriptaddress,showpacs,floatfix,%
preprintnumbers, nofootinbib,hyperref]{revtex4} 
 
\usepackage{epsfig} 
\usepackage{bm} 
 
\long\def\ignore#1{} 
 
\usepackage[dvips]{color} 
\definecolor{Black}{named}{Black} 
\definecolor{Blue}{named}{Blue} 
\definecolor{Red}{named}{Red}

\newcommand{\I}{{\rm i}} 
 
\newcommand{\D}{{\rm d}}

\def\beq{\begin{equation}}
\def\eeq{\end{equation}}

\begin{document} 
 
\title{Collective neutrino oscillations in non-spherical geometry}  
 
\author{Basudeb Dasgupta} 
\affiliation{Tata Institute of Fundamental Research, Homi Bhabha 
Road, Mumbai 400005, India} 
 
\author{Amol Dighe} 
\affiliation{Tata Institute of Fundamental Research, Homi Bhabha 
Road, Mumbai 400005, India} 
 
\author{Alessandro Mirizzi} 
\affiliation{Max-Planck-Institut f\"ur Physik 
(Werner-Heisenberg-Institut), F\"ohringer Ring 6, 80805 M\"unchen, 
Germany} 
\affiliation{Istituto Nazionale di Fisica Nucleare, Roma,  Italy}
 
\author{Georg Raffelt} 
\affiliation{Max-Planck-Institut f\"ur Physik 
(Werner-Heisenberg-Institut), F\"ohringer Ring 6, 80805 M\"unchen, 
Germany} 
 
\date{21 May 2008} 
 
\preprint{MPP-2008-47, TIFR/TH/08-19} 

\begin{abstract} 
The rich phenomenology of collective neutrino oscillations has been 
studied only in one-dimen\-sional or spherically symmetric systems. 
Motivated by the non-spherical example of coalescing neutron stars, 
presumably the central engines of short gamma-ray bursts, we use the 
Liouville equation to formulate the problem for general source 
geometries.  Assuming the neutrino ensemble displays self-maintained 
coherence, the problem once more becomes effectively one-dimensional 
along the streamlines of the overall neutrino flux. This approach 
for the first time provides a formal definition of the ``single-angle 
approximation'' frequently used for supernova neutrinos and allows 
for a natural generalization to non-spherical geometries. We study 
the explicit example of a disk-shaped source as a proxy for 
coalescing neutron~stars. 
\end{abstract} 
 
 
\maketitle 
 
\section{Introduction}                        \label{sec:introduction} 
 
Flavor transformations caused by neutrino mixing depend on the matter 
background and on the neutrino fluxes themselves: neutrino-neutrino 
interactions provide a nonlinear term in the equations of 
motion~\cite{Pantaleone:1992eq, Sigl:1992fn} that gives rise to 
collective flavor transformation phenomena~\cite{Samuel:1993uw, 
Kostelecky:1993dm, Kostelecky:1995dt, 
  Samuel:1996ri, Pastor:2001iu, Wong:2002fa, Abazajian:2002qx, 
  Pastor:2002we, Sawyer:2004ai, Sawyer:2005jk, Sawyer:2008zs}. 
The neutrino density needs to be so large that a typical 
neutrino-neutrino interaction energy $\mu$ is comparable to the 
vacuum oscillation frequency $\omega=\Delta m^2/2E$. Only recently 
has it been fully appreciated that this condition is sufficient even 
if a dense background of ordinary matter provides a much larger 
interaction energy so that naively neutrino-neutrino interactions 
would seem negligible~\cite{Duan:2005cp, 
  Duan:2006an}. Following this crucial insight, nonlinear 
oscillation phenomena in the supernova (SN) context have been studied 
over the past two years in a long series of papers~\cite{Duan:2006an, 
  Hannestad:2006nj, Duan:2007mv, Raffelt:2007yz, EstebanPretel:2007ec, 
  Raffelt:2007cb, Raffelt:2007xt, Duan:2007fw, Fogli:2007bk, 
  Duan:2007bt, Duan:2007sh, Dasgupta:2008cd, EstebanPretel:2007yq, 
  Dasgupta07, Duan:2008za, Dasgupta:2008my, Duan:2008eb}. 
 
One striking effect is ``self-maintained 
coherence''~\cite{Samuel:1993uw, Kostelecky:1993dm, 
Kostelecky:1995dt, Samuel:1996ri}. Different neutrino modes have 
different vacuum oscillation frequencies $\omega=\Delta m^2/2E$, but 
with strong neutrino-neutrino interactions they ``stick together'' 
and oscillate as a single mode characterized by the ``synchronized 
oscillation frequency'' $\omega_{\rm sync}=\langle\Delta 
m^2/2E\rangle$. This can lead to all modes going through an MSW 
resonance together, the ``collective MSW-like transition'' or 
``synchronized MSW effect''~\cite{Wong:2002fa, Abazajian:2002qx, 
Duan:2007fw, Duan:2007sh, Dasgupta:2008cd}. 
 
More interesting still are collective phenomena driven by the 
decrease of the neutrino flux with distance from the source. The 
adiabatic transition from a dense to a dilute neutrino gas produces 
step-like spectral features where the spectrum sharply splits into 
parts of different flavor transformation, so-called ``step-wise 
spectral swapping'' or ``spectral splits''~\cite{Duan:2006an, 
Raffelt:2007cb, Raffelt:2007xt, 
  Duan:2007fw, Duan:2007bt, Fogli:2007bk, Duan:2007sh, 
  Dasgupta:2008cd, Duan:2008za}. Spectral splits can result from 
a preceding collective MSW effect (``MSW prepared spectral split'') 
or from neutrino-neutrino interactions alone. 
 
The latter case depends on an unusual form of non-equilibrium among 
neutrino flavors where one has an excess of  flavor pairs, say
$\nu_e\bar\nu_e$, over the other flavors. 
For neutrinos streaming off a SN core one 
indeed expects a hierarchy of number fluxes 
$F_{\nu_e}>F_{\bar\nu_e}>F_{\nu_\mu,\nu_\tau}=F_{\bar\nu_\mu,\bar\nu_\tau}$.
Therefore, one can have collective transformations of the form 
$\nu_e\bar\nu_e\to \nu_x\bar\nu_x$,
where $x$ stands for some suitable combination of 
$\mu$ and $\tau$ neutrinos. 
These ``collective pair transformations'' do not 
violate any conservation law and thus can be catalyzed even by a very 
small mixing angle. $F_{\bar\nu_e}$ can completely swap with 
$F_{\bar\nu_x}$ whereas the larger $F_{\nu_e}$ converts only to the 
extent allowed by flavor-lepton conservation, but in a step-like 
spectral form. Both the complete conversion of 
$F_{\bar\nu_e}$ and the split in 
the $\nu_e$ spectrum provide signatures for the inverted neutrino 
hierarchy even for an extremely small 13-mixing 
angle~\cite{Duan:2007bt, Dasgupta:2008my}.

For non-isotropic enviroments,
multi-angle effects may play an important role.
The term ``multi-angle effects'' actually refers to two different 
issues. One is that the weak interaction 
potential between two relativistic particles is proportional to 
($1-\cos\theta$), where $\theta$ is their relative angle of propagation. 
One usually considers an isotropic background of ordinary matter so 
that $\cos\theta$ averages to zero. The same is true in an isotropic 
neutrino gas for the neutrino-neutrino term. 
The second issue is the ``multi-angle instability''.
Neutrinos arriving from different points on the source 
belong to different angular modes, which
may decohere kinematically in flavor 
space~\cite{Sawyer:2004ai, Sawyer:2005jk, Sawyer:2008zs, 
Raffelt:2007yz, EstebanPretel:2007ec}. This effect can be 
self-induced in the sense that a very small initial anisotropy 
is enough to 
trigger an exponential runaway, for example in a gas consisting of 
equal densities of neutrinos and antineutrinos~\cite{Raffelt:2007yz}. 
Systems consisting of very few angular modes can show a two-stream or 
multi-stream instability~\cite{Sawyer:2004ai, Sawyer:2005jk, 
Sawyer:2008zs}. On the other hand, numerical studies show that 
systems consisting of many angular modes and with a sufficient 
neutrino-antineutrino asymmetry do not show a multi-angle instability 
but rather show self-maintained coherence among different angular 
modes~\cite{Duan:2006an,EstebanPretel:2007ec,Fogli:2007bk}. The SN neutrino flux parameters 
seem to be such that the multi-angle instability plays no role in 
practice. Based on this assumption, most of the SN studies have used the 
``single-angle approximation'', where all angular modes are assumed to 
have the same behavior. 
 
Multi-angle effects related to the ($1-\cos\theta$) structure of the 
neutrino-neutrino term are unavoidable for an extended source radiating 
neutrinos into space because the emitted neutrino flux cannot form 
an isotropic gas. However, these effects also occur, and are 
easier to study theoretically, in a homogeneous system evolving in 
time that has a non-isotropic angular distribution of neutrinos.

In practice, however, one usually deals with stationary systems where 
one asks for the spatial variation of a neutrino ensemble as a 
function of distance from the source. Even if the neutrino-neutrino 
interaction were isotropic, we still would have geometric multi-angle 
effects because neutrinos reaching a certain point from an extended 
source have traveled on different trajectories. Even the simple case 
of an infinite radiating plane is not trivial. 
Here the direction perpendicular to the 
plane is the only direction in which the overall neutrino ensemble can 
show any spatial variation.  
Even if all neutrinos have 
the same energy and thus oscillate with the same frequency along their 
trajectories, the projection on the direction perpendicular to the 
plane yields different effective oscillation frequencies and thus 
kinematical decoherence. 
Neutrino-neutrino effects can synchronize 
different angular modes so that a sufficiently dense neutrino gas will 
not show this form of multi-angle decoherence. 
On the contrary, all 
angular modes will vary with the same oscillation length as a function 
of distance from the plane. A similar description applies to a 
spherical source where one asks for the variation of all angular modes 
along the radial direction. 
 
The single-angle treatment of SN neutrino oscillations amounts to the 
assumption of self-maintained coherence among angular modes, although 
it has never been explicitly expressed in this form. 
Some authors assumed that 
all angular modes oscillate as the radial 
one~\cite{Duan:2006an}. However, in this case the neutrino-neutrino 
interaction vanishes because of the ($1-\cos\theta$) factor,
so it was necessary to assume a certain average of the neutrino-neutrino 
interaction strength. Other authors represented all angular modes by a 
single angular mode radiated at $45^\circ$ relative to the radial 
direction and then used a neutrino-neutrino interaction strength 
consistent with this assumption~\cite{EstebanPretel:2007yq}.  This 
implementation of the single-angle approximation has the advantage 
that one can use the same numerical code as for multi-angle 
simulations, simply restricting oneself to a single angular bin. In 
yet other cases the system was modeled as a homogeneous and isotropic 
gas that evolves in time, assuming a time variation of the neutrino 
density that mimics the radial variation in the spherical case. 
 
One of our goals is to show that the single-angle treatment can be 
formulated self-consistently. The assumption that all angular modes 
evolve the same in flavor space provides a unique concept of what is 
meant by ``single-angle behavior.'' This is straightforward in the 
systems described so far where symmetry dictates that the spatial 
variation is only along a certain direction, effectively reducing the 
problem to one dimension. 
We are really motivated, however, by more general geometries where no 
special direction is singled out by symmetry. In particular, we are 
interested in the case of coalescing neutron stars that may form the 
inner engines of short gamma-ray bursts~\cite{Ruffert:1998qg}.

The accretion torus or disk formed during neutron star
coalescence is a neutrino source comparable to a SN core. 
However, the torus is less dense and not efficient at producing 
$\nu_\mu$ and $\nu_\tau$. Therefore, the torus is a source for a 
dominant $\nu_e\bar\nu_e$ pair flux which is thought to produce an 
$e^+e^-$ pair plasma, thus powering short gamma-ray bursts. 
The annihilation cross section for $\nu_e\bar\nu_e\to e^+e^-$ is
much larger than that for $\nu_x\bar\nu_x\to e^+e^-$, 
so the neutrino flavor composition strongly influences the
number of $e^+ e^-$ pairs produced.
Therefore, one may ask if collective pair conversions occur 
in this environment close enough to the source to modify 
the energy transfer to the $e^+ e^-$ plasma,
and hence affect the strength of the gamma ray burst.

For coalescing neutron stars  
one expects a flux hierarchy $F_{\bar\nu_e}>F_{\nu_e}\gg 
F_{\nu_x}=F_{\bar\nu_x}$, which  differs from the SN case because 
the matter leptonizes when neutrons convert to protons, in contrast 
to the deleptonization of a SN core. The asymmetry between 
$F_{\nu_e}$ and $F_{\bar\nu_e}$  could be enough to prevent 
multi-angle decoherence so that similar collective effects as in the 
SN environment are conceivable. 
However, even 
granting this assumption, it is not straightforward how to 
implement something like a single-angle approximation in this context 
because it is not obvious how one should picture self-maintained 
coherence.
 
The purpose of our paper is to formulate the meaning of 
self-maintained coherence 
for general source geometries and study its implications. 
We find that the flavor 
variation  reduces to a quasi one-dimensional 
problem along the streamlines of the total neutrino flux. The main 
difference between the general case and the radiating plane or sphere 
is that the streamlines are typically curved, at least close to the 
source, so that self-maintained coherence applies to flavor 
oscillations along these curved streamlines. 
 
We begin in Sec.~\ref{sec:EOMs} with the general equations of motion 
for the neutrino matrices in flavor space. In 
Sec.~\ref{sec:synchronized oscillations} we formulate the collective 
equations for neutrinos only (no antineutrinos) and consider only 
synchronized oscillations.
Sec.~\ref{sec:streamlines} shows the existence of streamlines
and gives a prescription for calculating the flavor evolution
along them. In Sec.~\ref{sec:geometries} we solve the problem for several 
geometries. In Sec.~\ref{sec:bipolar} we study the generalization to a 
mixed system of neutrinos and antineutrinos where collective pair 
transformations are possible. In Sec.~\ref{sec:neutron stars} we 
consider an explicit example for a disk source with parameters 
inspired by numerical simulations of coalescing neutron stars.  We 
summarize our conclusions in Sec.~\ref{sec:conclusions}.

\section{Formalism for an arbitrary source geometry}
\label{sec:EOMs} 

\subsection{General framework}
 
A homogeneous ensemble of mixed neutrinos can be described by 
``matrices of density'' $\varrho_{\bf p}$, which really are matrices 
of occupation numbers, for each momentum mode ${\bf p}$ 
\cite{Dolgov:1980cq, Rudzsky:1990, Sigl:1992fn, mckellar&thomson}. 
If 
$a^\dagger_{i,{\bf p}}$ and $a_{i,{\bf p}}$ are the creation and 
annihilation operators of a neutrino in the mass eigenstate $i$ of 
momentum ${\bf p}$, we have $(\varrho_{\bf p})_{ij}\propto\langle 
a^\dagger_{j} a_{i}\rangle_{\bf p}$ so that the diagonal entries of 
$\varrho_{\bf p}$ are the usual occupation numbers (expectation 
values of number operators), whereas the off-diagonal elements encode 
the phase relations that allow one to follow flavor oscillations. 
Such a description assumes that higher-order correlations beyond 
field bilinears play no role, probably a good approximation for 
neutrinos produced from essentially thermal sources such as the 
early-universe plasma or a SN core. 
 
Antineutrinos are described in an analogous way by $(\bar\varrho_{\bf
  p})_{ij}=\langle\bar a^\dagger_{i} \bar a_{j}\rangle_{\bf p}$. Note
that we always use overbars to characterize antiparticle quantities.
The order of flavor indices was deliberately interchanged on the
r.h.s.\ so that the matrices $\varrho_{\bf p}$ and $\bar\varrho_{\bf
  p}$ transform in the same way in flavor space~\cite{Sigl:1992fn}. In
this way one can, for example, write the overall neutrino current in
the simple form $\int \D{\bf p}\,v_{\bf p} (\varrho_{\bf
  p}-\bar\varrho_{\bf p})$ where here and henceforth $\D{\bf p}$
stands for $\D^3{\bf p}/(2\pi)^3$. While some authors prefer the
seemingly more intuitive equal order of flavor indices for neutrinos
and antineutrinos, in that convention the current would be $\int
\D{\bf p}\,v_{\bf p} (\varrho_{\bf p}-\bar\varrho_{\bf p}^*)$ and
generally the equations will involve both the matrices of density and
their complex conjugates. (In a truly
field-theoretic derivation there is no ambiguity about the relative
structure of neutrino and antineutrino matrices.  An analogous example
for matrices of occupation numbers is provided by the kinetic
treatment of the quark--gluon plasma where the quark distribution
functions are $3\times3$ matrices in color
space~\cite{Mrowczynski:2007hb}. The matrices for quarks and the ones
for antiquarks transform equally under a color gauge transformation.)
 
The matrices $\varrho_{\bf p}$ and $\bar\varrho_{\bf p}$  depend 
on time. Their evolution is governed by the Boltzmann collision 
equation 
\begin{equation} 
 \partial_t\varrho_{\bf p}= 
 -\I[{\sf\Omega}_{\bf p},\varrho_{\bf p}] 
 +\dot\varrho_{\bf p}\big|_{\rm coll} 
\end{equation} 
and a similar equation for $\bar\varrho_{\bf p}$. 
The collision term 
will be of no further concern in our paper because we only study 
freely streaming neutrinos. Further, $[\cdot,\cdot]$ is a commutator 
and ${\sf \Omega}_{\bf p}$ is the matrix of oscillation frequencies. 
In vacuum we have ${\sf\Omega}_{\rm p}={\sf M}^2/2|{\bf p}|$ with 
${\sf M}$ the neutrino mass matrix. In general ${\sf\Omega}_{\rm p}$ 
also depends on the background medium and notably on the presence of 
other neutrinos. Here and henceforth the ultrarelativistic limit for 
neutrinos is assumed. In other words, it is assumed that the 
difference between the neutrino energy $E_{\bf p}$ and momentum 
$p=|{\bf p}|$ is irrelevant except for oscillations: the only 
relevant difference between energy and momentum is captured by the 
matrix of oscillation frequencies. 
 
Up to this point we have considered a homogeneous ensemble evolving 
in time, largely relevant for neutrino oscillations in the early 
universe for which this formalism was originally developed. 
However, for a realistic representation of radiating 
objects such as supernovae or coalescing neutron stars we need to 
include spatial transport phenomena. To this end 
one introduces space-dependent occupation numbers or Wigner 
functions $\varrho_{\bf p,x}$. The quantum-mechanical uncertainty 
between location and momentum implies that this 
construction is only valid in the limit where spatial variations of 
the ensemble are weak on the microscopic length scales defined by the 
particles' typical Compton wavelengths. The left hand side 
of the Boltzmann collision equation now turns into the usual Liouville 
operator~\cite{Strack:2005ux} 
\begin{equation} 
 \partial_t\varrho_{\bf p,x} 
 +{\bf v}_{\bf p}\cdot{\nabla}_{\bf x}\,\varrho_{\bf p,x} 
 +\dot{\bf p}\cdot{\nabla}_{\bf p}\,\varrho_{\bf p,x}. 
\end{equation} 
The transition to the Liouville operator may seem obvious, but making 
it conceptually precise requires a long 
argument~\cite{Cardall:2007zw}.  The first term represents an explicit 
time dependence, the second a drift term caused by the particles' free 
streaming, and the third the effect of external macroscopic forces, 
for example gravitational deflection. Henceforth we shall neglect 
external forces and only retain the drift term. 

In our 
ultrarelativistic approximation the modulus of ${\bf v}_{\bf p}$ is 
the speed of light. Of course, for propagation over very large 
distances this may be a bad approximation when time-of-flight effects 
play a role. In this case the drift term would have a more complicated 
structure because the velocity is then also a nontrivial matrix in 
flavor space~\cite{Sigl:1992fn}. 
 
We shall focus on stationary problems. Even neutrino emission from a 
SN or coalescing neutron stars fall in this category unless there are 
fast variations of the source. In other words, we shall ignore a 
possible explicit time dependence of $\varrho_{\bf p,x}$ so that 
finally we arrive~at 
\begin{equation}\label{eq:EOM1} 
 {\bf v}_{\bf p}\cdot{\bf\nabla}_{\bf x}\,\varrho_{\bf p,x} 
 =-\I[{\sf\Omega}_{\bf p,x},\varrho_{\bf p,x}]\,, 
\end{equation} 
where the matrix of oscillation frequencies in general also depends 
on space because of the influence of matter and other neutrinos. 
 
Equation~(\ref{eq:EOM1}) could have been guessed starting from the 
usual treatment of neutrino oscillations. The neutrino density matrix 
describing a stationary beam in the $z$--direction follows 
$\partial_z\varrho_{E,z}=-\I[\Omega_{E,z},\varrho_{E,z}]$.  Moreover, 
in our ultrarelativistic approximation $E=p$ and $|{\bf v}_{\bf 
  p}|=1$. Equation~(\ref{eq:EOM1}) then simply amounts to a collection 
of many beams in different directions originating from different 
source points. 
 
We finally spell out Eq.~(\ref{eq:EOM1}) explicitly under the 
assumption that the only nontrivial ingredients consist of 
neutrino-neutrino refractive effects, 
\begin{widetext} 
\begin{eqnarray}\label{eq:EOM2} 
 \I\,{\bf v}_{\bf p}\cdot{\bf\nabla}_{\bf x}\,\varrho_{\bf p,x} 
 &=&+\frac{1}{2p}\,\bigl[{\sf M}^2,\varrho_{\bf p,x}\bigr] 
 +\sqrt{2}\,G_{\rm F} 
 \int \D{\bf q}\,(1-{\bf v}_{\bf q}\cdot{\bf v}_{\bf p}) 
 \bigl[(\varrho_{\bf q,x}-\bar\varrho_{\bf q,x}), 
 \varrho_{\bf p,x}\bigr]\,, \nonumber
 \\* 
 \I\,{\bf v}_{\bf p}\cdot{\bf\nabla}_{\bf x}\,\bar\varrho_{\bf p,x} 
 &=&-\frac{1}{2p}\,\bigl[{\sf M}^2,\bar\varrho_{\bf p,x}\bigr] 
 +\sqrt{2}\,G_{\rm F} 
 \int \D{\bf q}\,(1-{\bf v}_{\bf q}\cdot{\bf v}_{\bf p}) 
 \bigl[(\varrho_{\bf q,x}-\bar\varrho_{\bf q,x}), 
 \bar\varrho_{\bf p,x}\bigr]\,. 
\end{eqnarray} 
\end{widetext} 
The only difference between the neutrino and antineutrino equations 
is the sign of the vacuum oscillation term. This structure is a 
consequence of defining the antineutrino matrices with reversed 
flavor indices.

\subsection{EOMs with self-maintained coherence} 
\label{sec:synchronized oscillations} 
 
The nonlinear equations of motion (EOMs) Eq.~(\ref{eq:EOM2}) simplify 
considerably if self-maintained coherence occurs in a dense neutrino 
gas and all modes can be assumed to evolve in the same way.  For the 
moment we restrict ourselves to a source radiating only neutrinos (no 
antineutrinos). What we mean by self-maintained coherence in this 
case is that at a given location all neutrino modes are aligned with 
each other, assuming they were aligned at the source.
That is, 
\begin{equation} 
\varrho_{\bf p,x}={\sf P}_{\bf x}\,f_{\bf p,x}\,. 
\end{equation} 
Here, $f_{\bf p,x}={\rm Tr}(\varrho_{\bf p,x})$ is a scalar 
occupation number, summed over all flavors, while for $N$ flavors 
${\sf P}_{\bf x}$ is a $N\times N$ matrix normalized as ${\rm 
Tr}({\sf P}_{\bf x})=1$. We also define the local neutrino density 
and flux as 
\begin{eqnarray} 
n_{\bf x}&=&\int\D{\bf p}\,f_{\bf p,x}\,, 
\nonumber\\ 
{\bf F}_{\bf x}&=&\int\D{\bf p}\,{\bf v}_{\bf p}\,f_{\bf p,x}\,
= \langle {\bf v} \rangle_{\bf x} n_{\bf x} \; ,
\end{eqnarray} 
where 
\beq
\langle Q \rangle _{\bf x} 
\equiv \frac{ \int\D{\bf p}\, Q _{\bf p, x} \, f_{\bf p, x}}{
\int\D{\bf p} \, f_{\bf p,x}}
\eeq
is the momentum average of a quantity
$Q$ at location ${\bf x}$ with
respect to the distribution function $f_{\bf p, x}$.
For future convenience we also define
\beq
\widehat{\bf F}_{\bf x} \equiv 
\frac{{\bf F}_{\bf x}}{|{\bf F}_{\bf x}|} 
= \frac{\langle {\bf v} \rangle _{\bf x}}{
|\langle {\bf v} \rangle_{\bf x}|} 
\; ,
\eeq
a unit vector along the direction of flux ${\bf F}_{\bf x}$, or
equivalently, along the average velocity 
$\langle {\bf v} \rangle _{\bf x}$.

In this language, the EOM Eq.~(\ref{eq:EOM2}) becomes 
\begin{equation}\label{eq:EOM4} 
 {\bf v}_{\bf p}\cdot{\bf\nabla}_{\bf x} 
 \,f_{\bf p,x} 
 {\sf P}_{\bf x} 
 =-\frac{\I}{2p}\,\bigl[{\sf M}^2,{\sf P}_{\bf x}\bigr] 
 \,f_{\bf p,x}\,, 
\end{equation} 
where the nonlinear term has disappeared because it involves the 
commutator $[{\sf P}_{\bf x},{\sf P}_{\bf x}]=0$. The only impact of 
the neutrino-neutrino term is to glue the different modes together, 
but afterwards it no longer appears in the equation. 
 
The evolution now factorizes into the free streaming of the overall 
neutrino flux and the oscillation of the common flavor matrix ${\sf 
P}_{\bf x}$. The trace of Eq.~(\ref{eq:EOM4}) provides 
\begin{equation} 
 {\bf v}_{\bf p}\cdot{\bf\nabla}_{\bf x}\,f_{\bf p,x}=0\,, 
\label{eq:vpdelf}
\end{equation} 
allowing us to determine the overall neutrino density $n_{\bf x}$ and 
flux ${\bf F}_{\bf x}$ at any point, assuming it is given at the 
source surface. With ${\bf v}_{\bf p}\cdot{\bf\nabla}_{\bf 
x}\,f_{\bf p,x} ={\bf\nabla}_{\bf x}\cdot{\bf v}_{\bf p}\,f_{\bf 
p,x}$, integrating Eq.~(\ref{eq:vpdelf}) over all modes yields 
\begin{equation}\label{eq:EOM4a} 
 {\bf\nabla}_{\bf x}\cdot{\bf F}_{\bf x}=0\,, 
\end{equation} 
expressing the absence of neutrino sources or sinks. 
 
The EOM for ${\sf P}_{\bf x}$ is found by integrating 
Eq.~(\ref{eq:EOM4}) over all modes, 
\begin{equation}\label{eq:EOM4b} 
 {\bf\nabla}_{\bf x}\cdot 
 \bigl({\bf F}_{\bf x}\,{\sf P}_{\bf x}\bigr) 
 =-\frac{\I}{2}\,\bigl[{\sf M}^2,{\sf P}_{\bf x}\bigr] 
 \,\int\D{\bf p}\,\frac{f_{\bf p,x}}{p}\,. 
\end{equation} 
The l.h.s.\ expands as $\bigl({\bf\nabla}_{\bf x}\cdot{\bf F}_{\bf 
x}\bigr)\,{\sf P}_{\bf x} +{\bf F}_{\bf x}\cdot\bigl({\bf\nabla}_{\bf 
x}{\sf P}_{\bf x}\bigr)$ where the first term disappears because of 
Eq.~(\ref{eq:EOM4a}) so that 
\begin{equation}\label{eq:EOM4c} 
 {\bf F}_{\bf x}\cdot{\bf\nabla}_{\bf x}{\sf P}_{\bf x} 
 =-\frac{\I}{2}\,\bigl[{\sf M}^2,{\sf P}_{\bf x}\bigr] 
 \,\int\D{\bf p}\,\frac{f_{\bf p,x}}{p}\,. 
\end{equation} 
Dividing both sides by $|{\bf F}_{\bf x}|$, we get
\begin{equation}\label{eq:EOM4cc} 
 \widehat{\bf F}_{\bf x}\cdot{\bf\nabla}_{\bf x}{\sf P}_{\bf x} 
 =-\I\,\bigl[{\sf\Omega}_{\bf x},{\sf P}_{\bf x}\bigr]\, ,
\end{equation} 
where
\begin{equation}\label{eq:EOM4ccc} 
 {\sf\Omega}_{\bf x}=\frac{{\sf M}^2}{2}\, 
 \frac{\int\D{\bf p}\,p^{-1}\,f_{\bf p,x}} 
 {\left|\int\D{\bf p}\,{\bf v}_{\bf p}\,f_{\bf p,x}\right|} 
 =\frac{{\sf M}^2}{2}\, 
 \frac{\langle p^{-1}\rangle} 
 {|\langle{\bf v}\rangle_{\bf x}|}\, .
\end{equation} 
Equation~(\ref{eq:EOM4cc}) resembles the EOM in the single-angle 
approximation, with $\Omega_{\bf x}$ as the synchronized matrix 
of oscillation frequencies.
Note that $\langle p^{-1} \rangle$ is independent of location
if the energy spectrum of neutrinos is assumed to be
the same everywhere.

\subsection{Streamlines}
\label{sec:streamlines}

Equation~(\ref{eq:EOM4cc}) is a partial differential equation (PDE)
for the matrix ${\sf P}_{\bf x}$. 
It can be reduced to a set of ordinary differential equations (ODEs)
\begin{eqnarray}
\frac{\D {\bf x}}{\D s} & = & \widehat{\bf F}_{\bf x} \; , 
\label{ode1}  \\
\frac{\D {\sf P}_{\bf x}}{\D s} & = & 
-\I\,\bigl[{\sf\Omega}_{\bf x},{\sf P}_{\bf x}\bigr]\, ,
\label{ode2}
\end{eqnarray}
where $s$ is a parameter along the ``characteristic line,''
or ``streamline,'' defined by Eq.~(\ref{ode1}).
Since $\widehat{\bf F}_{\bf x}$ is unique at each location,
the streamlines do not intersect each other.
Along each streamline, the differential equation [Eq.~(\ref{ode2})] for
the matrix ${\sf P}_{\bf x}$ is a set of coupled, but
linear ODEs and hence can be solved
easily and uniquely, given the boundary conditions.

As an illustration, we now specialize to a two-flavor system,
where the coupled ODEs in Eq.~(\ref{ode2}) reduce to a 
single ODE and may be interpreted as the evolution of
a single phase.
As uaual, we express a Hermitian $2\times2$ matrix 
${\sf A}$ in terms of a polarization vector $\vec{\rm A}$ by 
virtue of ${\sf A}=\frac{1}{2}({\rm Tr}\,{\sf A} 
+\vec{A}\cdot\vec{\sigma})$ 
with $\vec{\sigma}$ the vector of Pauli matrices.\footnote{A 
sans-serif letter such as ${\sf A}$ indicates a matrix in flavor 
space, a letter with an arrow such as $\vec{A}$ indicates a vector in 
flavor space, and a bold-faced letter such as ${\bf A}$ a vector in 
coordinate space.}  The equation for synchronized oscillations then 
turns into the usual precession formula 
\begin{equation}\label{eq:EOM4f} 
 \widehat{\bf F}_{\bf x}\cdot{\bf\nabla}_{\bf x}\vec{P}_{\bf x} 
 =\omega_{\bf x}\,\vec{B}\times\vec{P}_{\bf x}\,, 
\label{precession}
\end{equation} 
where $\vec{P}_{\bf x}$ is the polarization vector corresponding to 
$\varrho_{\bf x}$ and $\omega_{\bf x} \vec{B}$ the one corresponding 
to $\sf\Omega_{\bf x}$.
Here $\vec{B}$ is a unit vector by definition,
which is determined by ${\sf M}^2$ and hence is independent of 
location.
The effective synchronized oscillation frequency is
\beq
\omega_{\bf x} = \frac{\Delta m^2}{2} 
\frac{\langle p^{-1} \rangle}{|\langle {\bf v} \rangle _{\bf x}|} 
= \frac{\omega_0}{|\langle {\bf v} \rangle_{\bf x}|} \; ,
\label{eq:omega0}
\eeq
where $\omega_0$ is the synchronized oscillation frequency
far away from the source, where the details of the source 
geometry do not matter.
All the effects of the source geometry are thus captured
by the quantity $|\langle {\bf v} \rangle_{\bf x}|$.

The problem becomes even simpler if we observe that 
the magnitude of $\vec{P}_{\bf x}$ and its
projection on $\vec{B}$ are conserved.
We can then write $\vec{P}_{\bf x}$ in terms of its components
along and perpendicular to $\vec{B}$ as 
\beq
\vec{P}_{\bf x} = (P_\perp \cos a_{\bf x}, P_\perp \sin a_{\bf x},
P_\parallel) \; .
\label{pvec-param}
\eeq
Equation~(\ref{eq:EOM4f}) then reduces to a scalar equation
\begin{equation}\label{eq:EOM4g} 
 \widehat{\bf F}_{\bf x}\cdot{\bf\nabla}_{\bf x}\,a_{\bf x} 
 = \omega_{\bf x} \; .
\end{equation} 
In other words, $\vec{P}_{\bf x}$ is fully specified by 
a real scalar number $a_{\bf x}$ that tells us the phase of 
the revolution of the polarization vector around $\vec{B}$. 
Equations~(\ref{ode1}) and (\ref{ode2}) in this case simplify to
\begin{eqnarray}
\frac{\D {\bf x}}{\D s} & = & \widehat{\bf F}_{\bf x}
= \frac{\langle {\bf v} \rangle_{\bf x}}{
|\langle {\bf v} \rangle_{\bf x}|} \; , 
\label{ode1:2flav} \\
\frac{\D a_{\bf x}}{\D s} & = & \omega_{\bf x} 
= \frac{\omega_0}{
|\langle {\bf v} \rangle_{\bf x}|} \; .
\label{ode2:2flav}
\end{eqnarray}
As before, the streamlines are given by Eq.~(\ref{ode1:2flav})
and the evolution is obtained by solving the single ODE
in Eq.~(\ref{ode2:2flav}).
This has been possible because of a simple parametrization
of $\vec{P}_{\bf x}$ in Eq.~(\ref{pvec-param}) in two
flavors.

In the case of three flavors, one can express a Hermitian $3\times 3$
matrix in terms of the Gell-Mann matrices~\cite{Dasgupta07}.
Equation~(\ref{precession}) then retains the same form, except that
the ``cross product'' is now understood to be $(\vec{A} \times
\vec{B})_i \equiv f_{ijk} A_j B_k$ with $f_{ijk}$ the structure
constants for SU(3).  The ODEs in Eq.~(\ref{ode2}) are coupled, but
linear, and can be solved along the streamlines.
 
The general strategy for calculating the neutrino flavor evolution is
therefore the following.  $(i)$ Find the streamlines for the problem
by solving Eq.~(\ref{ode1}) with the given neutrino source.  $(ii)$
Solve for the phase $a_{\bf x}$ (or for the matrix ${\sf P}_{\bf x}$)
by solving the ODEs [Eq.~(\ref{ode2})] along each streamline.  This
leads to the determination of $\varrho_{\bf p,x}$ everywhere.

An important consequence of the existence of streamlines is that the
evolution along any streamline is independent of the other
streamlines. Moreover, each streamline is determined solely by the
flux ${\bf F}_{\bf x}$ along it.  Thus the neutrino flavor content at
any point in space is determined simply by identifying the streamline
that passes through that point and evolving the phase $a_{\bf x}$
along the streamline. This effectively reduces any multi-dimensional
problem to one dimension.

\section{Flavor evolutions for various source geometries}
\label{sec:geometries}

We now calculate the flavor evolution of neutrinos in several physical
situations, ranging from the trivial to the complicated.  The crucial
quantity to calculate in each scenario is $\langle {\bf v} \rangle
_{\bf x}$, which captures the effect of the source geometry through
Eq.~(\ref{eq:omega0}).

\subsection{Infinite Plane} 
 
For a homogeneous plane that radiates neutrinos iso\-tropically, the 
streamlines are straight lines normal to the surface. 
Therefore, the oscillation phase $a$ depends only on the 
distance $z$ from the plane and the surfaces of equal phase 
are planes parallel to the radiating surface. 
The average velocity of all neutrinos away from the plane is half the 
speed of light, $\langle {\bf v} \rangle _{\bf x} =
\frac{1}{2} \hat{z}$ everywhere. Therefore, 
$\omega_{\bf x}=2\omega_0$ represents the flavor oscillations along 
the direction perpendicular to the plane. 
This gives $a = 2 \omega_0 \, z + a_0$, 
where $a_0$ may be chosen to be an arbitrary real number. 
This, along with the initial conditions, completely determines the 
flavor state of the neutrinos at any point in space.

\subsection{Sphere} 

For a spherical source, symmetry dictates that the streamlines are 
straight lines following the radial distance from the center and that 
the surfaces of equal phase are spherical shells around the source. 
Therefore spherical symmetry dictates that the average velocity 
$\langle {\bf v} \rangle _{\bf x}=\langle v_r \rangle \hat{r}$ 
is along the radius. 
All that remains to be determined is $\langle v_r \rangle$, 
the average radial velocity of all neutrinos at distance $r$. 

\begin{figure}[t] 
\begin{center} 
\epsfig{figure=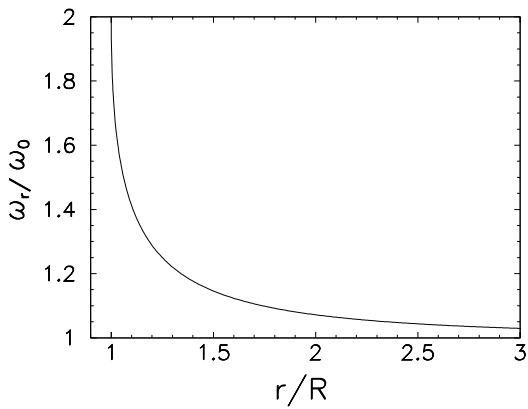,width=0.8\columnwidth,angle=0} 
\end{center} 
\caption{Synchronization frequency $\omega_{r}$ for an emitting 
  sphere (radius $R$) as a function of 
  distance~$r$.}  \label{fig:sphere} 
\end{figure}  
 
To this end we note that all modes that are radiated 
at the neutrino sphere with the same zenith angle $\vartheta_R$ 
relative to the radial direction behave identically.
Their radial velocity at the 
neutrino sphere is $v_R=\cos\vartheta_R$, so it is useful to 
classify all modes by their $v_R$.  Simple geometry tells us that 
$R\,\sin\vartheta_R=r\sin\vartheta_r$, where $\vartheta_r$ is the 
zenith angle of a given neutrino trajectory relative to the radial 
direction at a distance $r$.  Therefore, the radial velocity at 
distance $r$ of a mode that has radial velocity $v_R$ at the neutrino 
sphere is 
\begin{equation}\label{eq:vr} 
v_r=\cos \vartheta_r =\sqrt{1-\frac{R^2}{r^2}\,\left(1-v_R^2\right)}\;. 
\end{equation} 
 
Given the all-flavor phase-space density $f(v_R,R)$ of a mode $v_R$ 
at the neutrino sphere, we can determine its value $f(v_R,r)$ at 
another distance from flux conservation. In spherical coordinates we 
have $\partial_r[r^2 v_r f(v_R,r)]=0$ so that $f(v_R,r)\propto 
(R/r)^2 (v_R/v_r)$. 
 
If we assume isotropic emission at the neutrino sphere, the modes are 
uniformly distributed in the variable $0\leq\cos\vartheta_R\leq1$ or 
$0\leq v_R\leq1$, so that 
\begin{equation} 
\langle v_r\rangle= 
\frac{\int_0^1 \D v_R\, v_r\,f(v_R,r)}{\int_0^1 \D v_R\, f(v_R,r)} 
=\frac{\int_0^1 \D v_R\, v_R}{\int_0^1 \D v_R\,v_R/v_r}\,. 
\end{equation} 
Inserting the expression Eq.~(\ref{eq:vr}) for $v_r$ and 
performing the integral we find 
\begin{equation} 
\langle v_r\rangle=\frac{1}{2} 
\left[1+\sqrt{1-\left(\frac{R}{r}\right)^2}\,\right]\,. 
\end{equation} 
At the neutrino sphere ($r=R$) one finds $\langle v_R\rangle
=\frac{1}{2}$, identical to the radiating plane, whereas at large
distances $r\gg R$ all modes move essentially in the radial direction
and $\langle v_r\rangle\to 1$. Therefore, near the neutrino sphere,
the neutrino flavor evolves along the radial direction with frequency
$\omega_r=2\omega_0$, whereas at large distances it evolves with
$\omega_r=\omega_0$. This case is represented in
Fig.~\ref{fig:sphere}. The decrease of $\omega_r$ to its asymptotic
value is surprisingly fast. As usual, the phase is given by
$a=\int\omega_r dr + a_0$, and this along with initial conditions
specifies the flavour content of neutrinos all over space.

\begin{figure} 
\begin{center} 
\epsfig{figure=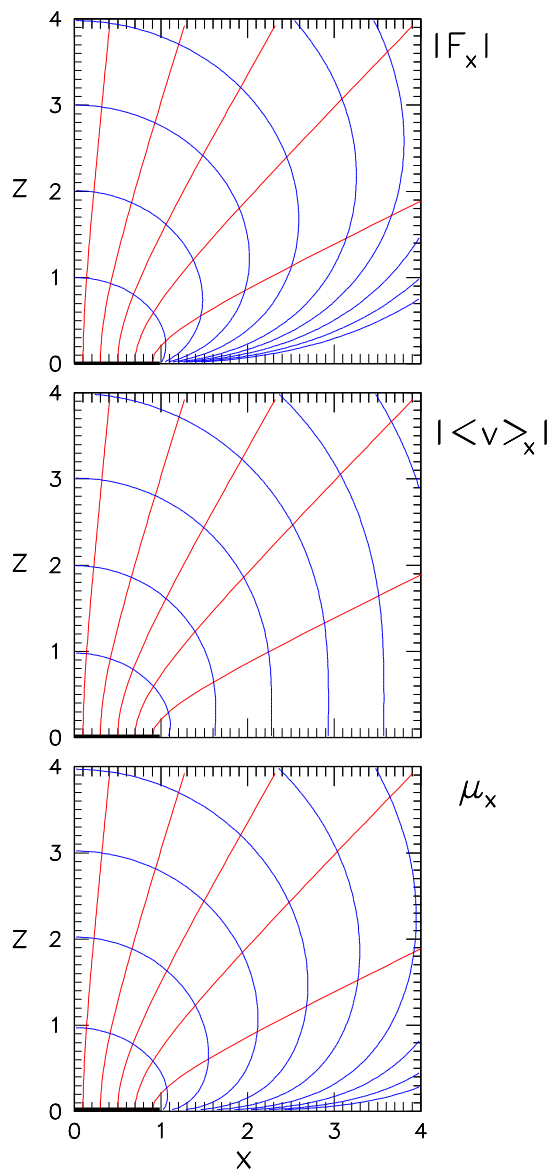,width =0.4\textwidth,angle=0} 
\end{center} 
\caption{Streamlines for the disk source, combined with contours for 
various quantities: (a)~Neutrino flux $|{\bf F}_{\bf x}|$. 
(b)~Average velocity $|\langle{\bf v}\rangle_{\bf x}|$. (c)~Effective 
neutrino-neutrino interaction strength $\mu_{\bf x}$.}\label{fig:streamlines} 
\end{figure} 

\begin{figure*} 
\begin{center} 
\epsfig{figure=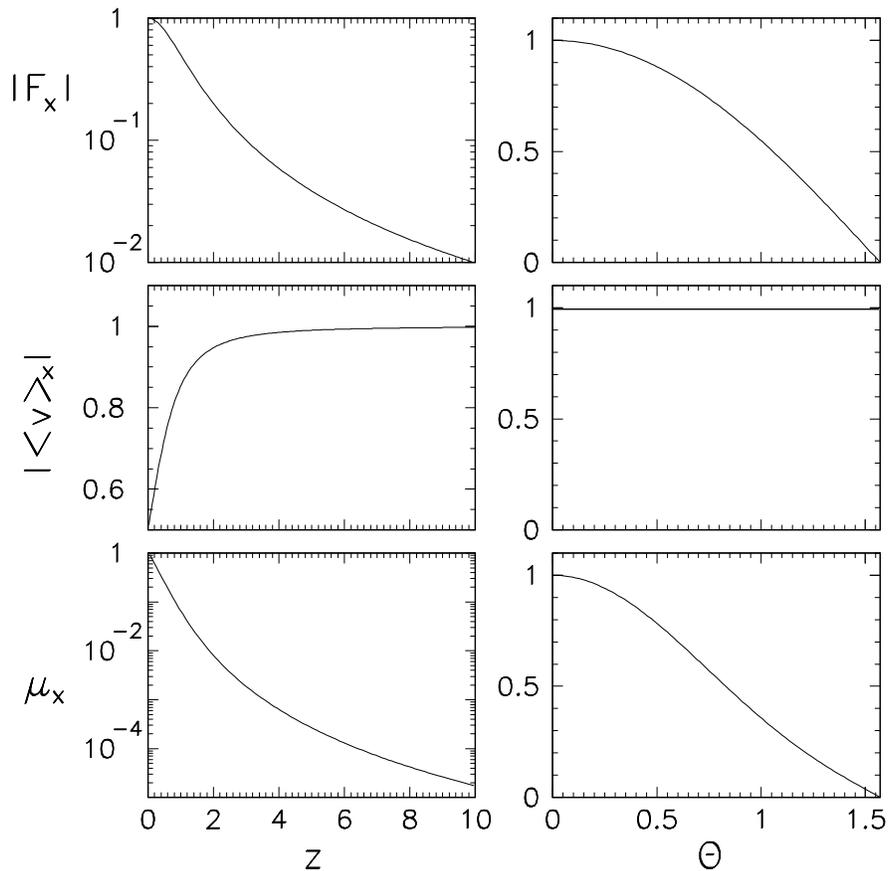,width =0.65\textwidth,angle=0} 
\end{center} 
\caption{Behavior as a function of $z$ for $\Theta=0$ (left panels) 
and as a function of $\Theta$ at infinite distance (right panels) of 
the following quantities for the disk source: (a)~Neutrino flux 
$|{\bf F}_{\bf x}|$. (b)~Average velocity $|\langle{\bf v}\rangle_
{\bf x}|$. 
(c)~Effective neutrino-neutrino interaction strength $\mu_{\bf x}$. 
\label{fig:asympt}} 
\end{figure*} 
 
\subsection{Disk}

Perhaps the simplest example where the streamlines are curved is the 
case of a radiating circular disk. This model approximately mimics 
the accretion disk formed by two coalescing neutron stars.  
The disk radius is taken to be $R$, setting the geometric scale 
for the system.  We assume that the disk radiates homogeneously a 
flux $F_0$ throughout its surface and that at 
each point the emission is identical to a blackbody 
source.

Very close to the disk the physical situation is identical to an
infinite radiating plane. The streamlines are perpendicular to the
disk with a flux $F_0$.  Viewed from a large distance, on the other
hand, the flux must be directed away from what looks like a point-like
source, so that the streamlines are radially outward from the
source. Since the source is assumed to behave like a ``blackbody'',
the neutrino flux coming from the source varies with $\cos\Theta$
where $\Theta$ is the zenith angle relative to the disk axis.  This is
simply because we are seeing a blackbody surface at an angle so that
the total flux is reduced by the projected size of the source. On the
other hand, the total flux passing through a spherical surface at
distance $r$ is conserved.  Therefore, at a distance $r\gg R$ we find
\begin{equation} 
|{\bf F}_{\bf x}|=F_0 (R/r)^2\,\cos\Theta\,. 
\end{equation} 
Therefore, we know the flux and its direction directly at the source 
and at a large distance.

We calculate the flux ${\bf F}_{\bf x}$ numerically 
and find that the streamlines are well represented by the
hyperbolae
\begin{equation} 
\frac{x^2}{\sin^2\Theta}-\frac{z^2}{\cos^2\Theta}=R^2\, .
\end{equation} 
Here we use coordinates $x$ along the disk radius and $z$ along the 
disk axis.
We have also calculated the magnitude of flux $|{\bf F}_{\bf x}|$, 
the magnitude of 
average velocity $|\langle {\bf v}\rangle_{\bf x}|=|{\bf F}_{\bf 
  x}|/n_{\bf x}$, and the effective neutrino-neutrino interaction 
strength $\mu_{\bf x}$ defined later in Eq.~(\ref{eq:mudefinition}). 
We show contours of these quantities in Fig.~\ref{fig:streamlines}. 
Moreover, in Fig.~\ref{fig:asympt} we show the behaviour of these 
quantities as a function of $z$ for $\Theta=0$ and their asymptotic 
variation as a function of $\Theta$ at a large distance from the 
disk. Details of the numerical calculations are given in 
Appendix~\ref{app:stream}.

\section{Neutrinos and Antineutrinos} 
\label{sec:bipolar} 
 
\subsection{Single Mode Equations} 
 
We now extend the previous considerations to the more interesting 
situation where both neutrinos and antineutrinos stream from the 
source. The main difference is that we now assume that neutrinos and 
antineutrinos each form two separate classes of modes that oscillate 
together, i.e., we assume that the behavior of all neutrino modes is 
well represented by a single matrix ${\sf P}_{\bf x}$ and that of the 
antineutrino modes by a matrix $\bar{\sf P}_{\bf x}$ such that 
$\varrho_{\bf p,x}={\sf P}_{\bf x} f_{\bf p,x}$ and $\bar\varrho_{\bf 
  p,x}=\bar{\sf P}_{\bf x} \bar f_{\bf p,x}$. 
 
Integrating the EOMs of Eq.~(\ref{eq:EOM2}) over all modes we 
arrive~at 
\begin{widetext} 
\begin{eqnarray}\label{eq:EOM5} 
 \I\,{\bf F}_{\bf x}\cdot{\bf\nabla}_{\bf x}\,{\sf P}_{\bf x} 
 &=&+\bigl[{\sf \Omega}_0,{\sf P}_{\bf x}\bigr]n_{\bf x} 
 +\sqrt{2}\,G_{\rm F}\left( 
 [(n_{\bf x}{\sf P}_{\bf x}-\bar n_{\bf x}\bar{\sf P}_{\bf x}), 
  n_{\bf x} {\sf P}_{\bf x}] 
 -[({\bf F}_{\bf x}{\sf P}_{\bf x}-\bar{\bf F}_{\bf x}\bar{\sf P}_{\bf x}), 
  {\bf F}_{\bf x} {\sf P}_{\bf x}]\right)\,, \nonumber \\ 
 \I\,\bar{\bf F}_{\bf x}\cdot{\bf\nabla}_{\bf x}\,\bar{\sf P}_{\bf x} 
 &=&-\bigl[\bar{\sf \Omega}_0,\bar{\sf P}_{\bf x}\bigr]\bar n_{\bf x} 
 +\sqrt{2}\,G_{\rm F}\left( 
 [(n_{\bf x}{\sf P}_{\bf x}-\bar n_{\bf x}\bar{\sf P}_{\bf x}), 
  \bar n_{\bf x} \bar{\sf P}_{\bf x}] 
 -[({\bf F}_{\bf x}{\sf P}_{\bf x}-\bar{\bf F}_{\bf x}\bar{\sf P}_{\bf x}), 
  \bar{\bf F}_{\bf x} \bar{\sf P}_{\bf x}]\right)\,. 
\end{eqnarray} 
\end{widetext} 
If it were not for the nonlinear term, of course, we would have 
synchronized oscillations separately for neutrinos and antineutrinos 
where ${\sf\Omega}_0$ is the oscillation matrix averaged over the 
neutrino spectrum whereas $\bar{\sf\Omega}_0$ is the one averaged 
over the antineutrino spectrum. 
 
In order to consider the simplest nontrivial case we assume further 
that the energy spectra for neutrinos and antineutrinos are the same 
as well as their angular emission characteristics. This implies 
$\Omega_0=\bar\Omega_0$. Moreover, we assume that the total flux of 
antineutrinos is $\alpha\leq 1$ times the flux of neutrinos. It 
simplifies the equations if we absorb this global factor in the 
normalization of the matrix $\bar{\sf P}_{\bf x}$ so that ${\rm 
Tr}(\bar{\sf P}_{\bf x})=\alpha$ whereas ${\rm Tr}({\sf P}_{\bf 
x})=1$. With these assumptions we have $\bar n_{\bf x}=n_{\bf x}$ and 
$\bar{\bf F}_{\bf x}={\bf F}_{\bf x}$. With $\langle{\bf 
v}\rangle_{\bf x}={\bf F}_{\bf x}/n_{\bf x}= \langle\bar{\bf 
v}\rangle_{\bf x}$ we then find 
\begin{eqnarray}\label{eq:EOM4aa} 
 \I\, \widehat{\bf F}_{\bf x}\cdot{\bf\nabla}_{\bf x}\,{\sf P}_{\bf x} 
 &=&+\bigl[{\sf \Omega}_{\bf x},{\sf P}_{\bf x}\bigr] 
 +\mu_{\bf x} 
 [({\sf P}_{\bf x}-\bar{\sf P}_{\bf x}),{\sf P}_{\bf x}]\,, \nonumber \\ 
 \I\, \widehat{\bf F}_{\bf x}\cdot{\bf\nabla}_{\bf x}\,\bar{\sf P}_{\bf x} 
 &=&-\bigl[{\sf \Omega}_{\bf x},\bar{\sf P}_{\bf x}\bigr] 
 +\mu_{\bf x} 
 [({\sf P}_{\bf x}-\bar{\sf P}_{\bf x}),\bar{\sf P}_{\bf x}]\,. 
\end{eqnarray} 
The oscillation matrix is ${\sf\Omega}_{\bf 
  x}={\sf\Omega}_0/|\langle{\bf v}\rangle_{\bf x}|$ whereas $\mu_{\bf 
  x}$ is the effective neutrino-neutrino interaction strength to be 
discussed below. 
 
If neutrinos and antineutrinos have the same streamlines, we find a 
unique extension of the idea of self-maintained coherence to a source 
with general geometry. Therefore, one may expect that the solutions 
along given streamlines depend on the source geometry primarily 
through the variation of $\mu$ along the streamlines. 
 
In the SN context, the evolution of the neutrino ensemble driven by 
the decrease of $\mu$ tends to be essentially adiabatic so that the 
exact variation of $\mu$ along a streamline should not matter much 
for the overall picture. This suggests that we may be allowed to 
borrow the insights gained in the SN context directly to more general 
cases. 
 
Even if neutrinos and antineutrinos do not follow exactly the same 
streamlines, if the lateral variation of the solution is smooth, 
one may speculate that the system finds some suitable average and 
that the final solution could still be similar to the simple case.

\subsection{Effective Neutrino-Neutrino Interaction Strength} 
 
In the limit of self-maintained coherence discussed in the previous 
section, we are led to define an effective neutrino-neutrino 
interaction strength 
\begin{eqnarray}\label{eq:mudefinition} 
\mu_{\bf x}&=&\sqrt2\,G_{\rm F}\,|{\bf F}_{\bf x}|\, 
\frac{n_{\bf x}^2-|{\bf F}_{\bf x}|^2}{|{\bf F}_{\bf x}|^2} 
\nonumber\\ 
&=&\sqrt2\,G_{\rm F}\,|{\bf F}_{\bf x}|\, 
\left(\frac{1}{|\langle{\bf v}\rangle_{\bf x}|^{2}}-1\right) 
\,. 
\end{eqnarray} 
In principle, the same coefficient would have appeared in the 
neutrino-only case, but there is was accompanied by a vanishing 
commutator so that it did not appear explicitly. The definition of 
$\mu_{\bf x}$ is unique except for the normalization that depends on our 
choice ${\rm Tr}({\sf P}_{\bf x})=1$ and ${\rm Tr}(\bar{\sf P}_{\bf 
  x})=\alpha$. In our normalization, $\mu_{\bf x}$ has the 
interpretation of the average neutrino-neutrino interaction energy and 
corresponds to the quantity $\sqrt 2 G_{\rm F}n_e$ in the ordinary 
matter effect. 
 
Equation~(\ref{eq:mudefinition}) implies some general features of 
the variation of $\mu$ at large distance $r$ from an extended but 
finite source. (We always consider sources that are not point like 
or else there would be no neutrino-neutrino effects, but that are 
finite, in contrast to the infinite plane mentioned earlier, or else 
there would be no systematic decrease of $\mu$ with distance.) The 
flux factor $|{\bf F}_{\bf x}|$ provides a trivial $r^{-2}$ scaling 
from the geometric flux dilution, as shown in 
Fig.~\ref{fig:asympt}. The ``collinearity factor'' 
\begin{equation} 
C_{\bf x}=\frac{1}{|\langle{\bf v}\rangle_{\bf x}|^{2}}-1 
\end{equation} 
arises from the $1-\cos\theta$ structure of the neutrino-neutrino 
interaction. At a large distance all neutrinos travel essentially 
collinear so that $|\langle{\bf v}\rangle_r|\to1$, implying an 
additional ``collinearity suppression.'' A typical transverse 
component of the neutrino velocity is of order $R/r$ if $R$ is the 
size of the source and $r$ the distance. Therefore, a typical neutrino 
velocity along the radial direction is $v_r\sim 1-(R/r)^2$, implying 
$C_r\propto r^{-2}$. Therefore, we find the general scaling 
\begin{equation} 
\mu_r\propto r^{-4} 
\end{equation} 
that is well known for spherical sources, and is also confirmed 
numerically in Fig.~\ref{fig:asympt}.

\begin{figure} 
\begin{center} 
\epsfig{figure=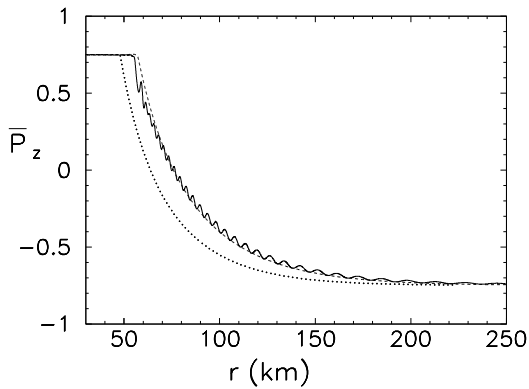,width =0.8\columnwidth,angle=0} 
\end{center} 
\caption{Radial evolution of the antineutrino polarization vector 
component  ${\bar P_z}$ for neutrinos emitted from a spherical 
source. We compare the multi-angle solution (continuous curve
that shows small oscillations) with the 
adiabatic limit of single-angle solutions where the dashed curve 
corresponds to Eq.~(\ref{eq:singleours}) and the dotted curve to 
Eq.~(\ref{eq:singlefull}).\label{fig:fig4}} 
\end{figure} 
 
For a spherical source we have derived an explicit expression for 
$\langle v_r\rangle$ in Eq.~(\ref{eq:vr}). In this case we find 
explicitly 
\begin{eqnarray} 
C_r&=&4\,\left[\frac{1-\sqrt{1-(R/r)^{2}}}{(R/r)^{2}}\right]^2 
-1 
\nonumber\\* 
&=&\frac{1}{2}\,\left(\frac{R}{r}\right)^2 
\quad\hbox{for $r\to\infty$\,.} 
\label{eq:singleours} 
\end{eqnarray} 
Immediately at the source we find $\mu_R=3$. The asymptotic behavior 
for large $r$ agrees with what one finds if one uses only a single 
angular bin with all neutrinos being radiated at $45^\circ$ relative 
to the radial direction~\cite{EstebanPretel:2007ec}. 
 
In the papers by Duan et al., beginning with Ref.~\cite{Duan:2005cp}, 
a somewhat different expression for $C_r$ was used because they 
pictured the single-angle approximation as all modes oscillating like 
the radial one. Considering the average neutrino-neutrino interaction 
they find the equivalent of 
\begin{eqnarray} 
C_r&=&\left[\frac{1-\sqrt{1-(R/r)^{2}}}{R/r}\right]^2 
\nonumber\\* 
&=&\frac{1}{4}\,\left(\frac{R}{r}\right)^2 
\quad\hbox{for $r\to\infty$\,.} 
\label{eq:singlefull} 
\end{eqnarray} 
While the scaling at large radius is, of course, the same, their 
normalization is a factor of two smaller. 
 
We have checked with a numerical multi-angle example that indeed our 
result provides the proper approximation. In Fig.~\ref{fig:fig4} we 
show the evolution of the antineutrino total polarization vector 
component ${\overline P}_z$ from a multi-angle simulation (continuous
curve). We compare it with the single-angle results corresponding to 
our approximation of $C_r$ of Eq.~(\ref{eq:singleours}) as a dashed
curve and with Duan et al.'s prescription of 
Eq.~(\ref{eq:singlefull}) as a dotted curve. In our numerical 
example, following Ref.~\cite{Hannestad:2006nj}, we have fixed the 
neutrino-neutrino interaction strength at the neutrino sphere 
(radius $R=10$~km) as $\mu_0=\sqrt{2}G_F F_0=0.7\times 
10^{5}$~km$^{-1}$, the neutrino oscillation frequency 
$\omega_0=0.3$~km$^{-1}$, the ratio of the number fluxes 
$\alpha=F_{\bar\nu_e}/F_{\nu_e}=0.75$, and the neutrino mixing angle 
$\theta_{13}=2.0\times 10^{-2}$. We have assumed the 
inverted mass hierarchy where  bipolar conversions  
occur. For the single-angle cases, we have used the adiabatic 
approximation worked out 
in~\cite{Duan:2007mv,Duan:2007fw,Raffelt:2007cb, 
Raffelt:2007xt},
where ${\overline P}_z$ depends on radius only through $\mu_r$ 
and where therefore there are no nutation wiggles. 
We discuss in more details about the adiabatic approximation below in 
Sec.~\ref{sec:adiab}. 
 
Of course, the difference found between Duan et al.'s approximation 
and ours is not crucial in that the neutrino fluxes and angular 
divergences assumed in previous studies had the character of toy 
examples. Since $\mu$ decreases with $r^{-4}$, a factor 2 
modification of $\mu$ translates into a 20\% radial shift of 
$\mu_r$ that is not important for a toy model. The difference close 
to the source is much larger, but less important. Close to the 
source we typically have strongly synchronized oscillations with a 
very small mixing angle, causing practically no difference in the 
overall solution. 
 
We finally consider the variation of $\mu$ for a disk-like source. 
At a large distance, the variation with distance will be 
proportional to $r^{-4}$ as in the case of a spherical source where 
a factor of $r^{-2}$ comes from the trivial geometric 
flux dilution with distance, while another factor $r^{-2}$ 
arises from the 
increasing collinearity of the neutrino trajectories with distance. 
To see this more explicitly we note that a typical ``radial'' 
velocity at a large distance is $v_r=(1-v_1^2-v_2^2)^{1/2}$ where 
$v_1$ and $v_2$ are typical transverse velocities in two orthogonal 
transverse directions. At a large distance we have $v_1\ll 1$ and 
$v_2\ll 1$ so that the factor $(1/|\langle {\bf v}\rangle_{\bf x}|^2-1)$
in Eq.~(\ref{eq:mudefinition}) is $\approx 
(v_1^2+v_2^2)$. A typical transverse velocity component of a 
neutrino at a distance $r$ is $R/r$ if $R$ is the geometric size of 
the source, confirming the $r^{-2}$ scaling of $C_r$ at a large 
distance.  

Moreover, we can also derive the variation of $\mu_{\bf x}$ 
with zenith angle $\Theta$ at a large distance. If the source is a 
circular disk of radius $R$, the azimuthal component of 
the velocity, $v_1$, will typically be $v_1 \propto (R/r)$.
The polar component $v_2$, on the other hand, 
will vary as $ v_2 \propto (R/r)\cos\Theta$, since the projected area
of the source in this direction is suppressed by $\cos\Theta$.
This leads to $C_r\propto (1+\cos^2\Theta)(R/r)^2$.  
Since the flux itself varies as $(R/r)^2\cos\Theta$ we expect 
\begin{equation} 
\mu_r\propto r^{-4}\,\cos\Theta\,(1+\cos^2\Theta)\,. 
\end{equation} 
This is numerically confirmed and agrees with the variation shown in 
Fig.~\ref{fig:asympt}. 
 
\subsection{Adiabatic approximation}             \label{sec:adiab} 

The adiabatic approximation for collective neutrino oscillations in
the single-angle case has been developed in a series of papers
~\cite{Duan:2007mv,Raffelt:2007cb, Raffelt:2007xt,Duan:2007fw}.  If
the neutrino flavor evolution is adiabatic, the flavor composition of
neutrinos as well as antineutrinos depends only on the initial state
and the value of $\omega_{\bf x}/\mu_{\bf x}$ at each ${\bf x}$, and
there is no need to compute the evolution explicitly.

Here, we follow~\cite{Raffelt:2007xt} and consider in particular the
case of a system constituted by only two polarization vectors,
$\vec{P}$ for neutrinos and $\vec{\overline{P}}$ for antinuetrinos.
At extremely large matter densities, the effective mixing angles are
small, as a result $\vec{P}$ and $\vec{\overline{P}}$ are aligned with
each other and with $\vec{B}$.  In the limit of a large but slow
varying neutrino interaction strength $\mu_{\bf x}$, the two
polarization vectors move in a pure precession mode around ${\vec B}$.
During the evolution, the conservation of ``flavor lepton number''
implies \beq P_z - \overline{P}_z = 1-\alpha \eeq is a conserved
quantity, where $\alpha$ is the ratio of antineutrino and neutrino
fluxes.

An explicit solution for the flavor evolution can be obtained simply
in terms of $\alpha$ as follows.  The results in
\cite{Raffelt:2007xt}, when calculated in the limit of vanishing
effective mixing angle, give 
\begin{eqnarray}\label{eq:adiab}
\frac{\omega_{\bf x}}{\mu_{\bf x}} &=&
\frac{1-\alpha+2\alpha\eta}{2}
-\frac{\eta\sqrt{\alpha\,(2-\alpha+\alpha\eta)}}{2\sqrt{1+\eta}}
\nonumber\\
&&-\frac{(1-\alpha+\alpha\eta)\sqrt{\alpha\,(1+\eta)}}{2\sqrt{2-\alpha+\alpha\eta}}
\end{eqnarray}
where $\eta$ is the cosine of the angle between
$\vec{\overline{P}}$ and $\vec{B}$ at any location.  It equals $+1$ to
begin with, and varies between $+1$ and $-1$ during the evolution.
Eq.~(\ref{eq:adiab}) can be inverted to yield $\eta$ at any location
as a function of $\omega_{\bf x}/\mu_{\bf x}$.  Knowing $\eta$, one
may compute the directions of $\vec{P}$ and $\vec{\overline{P}}$
through
\begin{eqnarray} 
{\overline P}_z &=& \alpha \eta \,\ , \nonumber \\ 
P_z &=& {\overline P}_z +(1-\alpha) \,\ .  
\label{eq:polarad} 
\end{eqnarray} 
The values of $P_z$ and $\overline{P}_z$ directly yield the
survival probabilities $P_{ee}$ and $P_{\bar{e} \bar{e}}$ of 
$\nu_e$ and $\bar{\nu}_e$ respectively.
$P_{ee}$ and $P_{\bar{e} \bar{e}}$ are thus direct functions of 
the local value of $\omega_{\bf x}/\mu_{\bf x}$.

In Fig.~\ref{fig:fig4} we have shown how the analytical adiabatic
approximation agrees very well with the multi-angle numerical
simulation for the spherically symmetric SN case. In general, this
result means that in the cases where neutrino ensembles exhibit a
self-maintained coherent behaviour, the adiabatic approximation
together with the single-angle limit described in
Eq.~(\ref{eq:mudefinition}) are sufficient to describe the collective
flavor evolution.

On the other extreme, in the case of ensembles showing 
kinematical decoherence, the adiabatic approximation is not
appropriate, and the final outcome is a complete equalization 
of the neutrino fluxes.

\section{Coalescing Neutron Stars}           \label{sec:neutron stars} 
 
\begin{figure}[b] 
\begin{center} 
\epsfig{figure=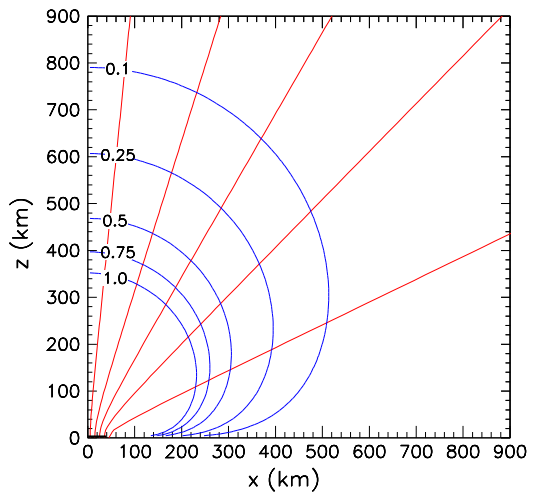,width =0.8\columnwidth,angle=0} 
\end{center} 
\caption{Streamlines and surfaces of constant $\omega_{\bf 
    x}/\mu_{\bf x}$ for neutrinos emitted by an accretion disk with 
  radius $R=50$~km. The surfaces are marked with the value of the 
  $\nu_e$ survival probability, assuming the adiabatic limit of 
  collective oscillations.\label{fig:fig5}} 
\end{figure} 
 
For purposes of illustration we now consider an explicit example for a 
disk-like source. Following the numerical studies of Ruffert and 
Janka~\cite{Ruffert:1998qg} for coalescing neutron stars, we use the 
following parameters for a disk source as an extremely schematic 
model. The all-species neutrino luminosity of the models shown by 
Ruffert and Janka varies between $(1-12) \times10^{52}~{\rm erg}~{\rm 
  s}^{-1}$ and the luminosity in species other than $\nu_e$ and 
$\bar\nu_e$ is always very small. Typically $\langle 
E_{\nu_e}\rangle<\langle E_{\bar\nu_e}\rangle$, a typical average 
being 12~MeV that we will use for both species. In contrast to the SN 
case, here one expects an excess of the ${\bar\nu}_e$ number flux over 
$\nu_e$, so that in this case we normalize the polarization vectors to 
the antineutrino flux. The ratio of number fluxes 
$\alpha=F_{\nu_e}/F_{\bar\nu_e}$ varies between about 0.65 and 0.81 if 
we ignore one case where the number fluxes are exactly equal and one 
where the $\nu_e$ flux practically vanishes. 
 
Inspired by these numbers and the pictures shown by Ruffert and Janka 
we adopt the following parameters for our homogeneous disk-like 
neutrino source: Radius~$R=50$~km, $L_{\bar\nu_e}=3\times10^{52}~{\rm 
  erg}~{\rm s}^{-1}$, $\langle E_{\nu}\rangle=12$~MeV, and 
$\alpha=0.75$. This implies a $\bar\nu_e$ flux at the disk of 
$F_0=2.0\times10^{43}~{\rm cm}^{-2}~{\rm s}^{-1}$ and a scale for the 
neutrino-neutrino interaction strength of $\mu_0=\sqrt{2}G_{\rm 
  F}F_0=8.4\times10^{-5}~{\rm eV}=4.3\times10^{5}~{\rm km}^{-1}$. We 
have in mind two-flavor oscillations driven by the atmospheric 
neutrino mass difference for which we use $\Delta m^2_{\rm 
  atm}=2.4\times10^{-3}~{\rm eV}^2$. Then the average vacuum 
oscillation frequency is $\omega_0=\langle\Delta m^2_{\rm 
  atm}/2E\rangle=1.50\times10^{-10}~{\rm eV}=0.76~{\rm km}^{-1}$. Here 
we have used that for an average neutrino energy $\langle 
E\rangle=12$~MeV and assuming a Maxwell-Boltzmann spectrum we have 
$\langle 1/E \rangle = 1/(8 ~{\rm MeV})$. 
 
The transition between the synchronized and the bipolar oscillation 
region, where the flavor pendulum begins to nutate and neutrino 
transformations begin in earnest, occurs at \cite{Hannestad:2006nj} 
\begin{equation} 
\frac{\omega_{\bf x}}{\mu_{\bf x}} \approx \frac{(1-\sqrt\alpha)^2}{2}\,, 
\label{borderline}
\end{equation} 
which is $9\times 10^{-3}$ for $\alpha=0.75$ in this scenario.
 
In Fig.~\ref{fig:fig5} we show the contours of constant $\omega_{\bf
  x}/\mu_{\bf x}$, which are the same as those of constant $P_{ee}$
under adiabatic approximation, for inverted hierarchy.  The contour
with $P_{ee}=1.0$ is the one where the condition of
Eq.~(\ref{borderline}) occurs and the synchronized to bipolar
transition takes place.  Within the region enclosed by this contour,
the oscillations are strongly synchronized so that macroscopically
nothing much happens. This contour delineates the area where the
flavor survival probability is essentially unity for both $\nu_e$ and
$\bar\nu_e$. Outside this region, neutrino transitions begin and we
show contours of the $\nu_e$ survival probability that indicates to
which extent the initial $\nu_e$ and $\bar\nu_e$ fluxes have been
converted to other flavors.

For our specific numerical example, we find that collective neutrino 
flavor conversions start at $z\agt 400$~km. Since the $\nu_e {\bar 
  \nu} _e$ annihilation rate per unit volume decreases very rapidly 
far from the source (for a spherical geometry one would expect $\sim 
r^{-8}$~\cite{goodman}) we do not expect that these flavor transitions 
affect significantly the neutrino energy deposition rate in the 
plasma.  However, the features of neutrino fluxes emitted from 
coalescing neutron stars are rather model dependent.  Here we 
speculate that if the asymmetry between the emitted neutrino species 
would be smaller, one could have collective flavor conversion or 
perhaps some sort of kinematical decoherence of the neutrino ensemble 
in a region close to the source, with a possible impact on the 
annihilation rate.  We also note that the matter density in the region 
close to the source is expected to be $\rho \sim 10^{10-12}~{\rm 
  g~cm}^{-3}$, thus it is so strong to prevent ordinary resonant 
flavor conversions~\cite{Volkas:1999gb}.  At larger distance one 
expects that two resonant level crossings will occur.  However, due to 
the uncertainties of the matter density profile in these regions, it 
is difficult to predict their effects on the neutrino burst. 
 
\section{Conclusions}                          \label{sec:conclusions} 
 
Neutrinos streaming from powerful astrophysical sour\-ces such as SN 
cores or coalescing neutron stars are so dense near the source that 
they must show nonlinear flavor oscillations induced by the 
neutrino-neutrino refractive effect. Numerical simulations reveal a 
rich variety of phenomena, some of which have been explained with 
simple analytic models. However, numerical simulations thus far have 
been restricted to homogeneous gases evolving in time or to sources 
with exact spherical symmetry. More general geometries are numerically 
much more demanding and have not yet been studied. 
 
Therefore, we have studied what might be expected under the 
assumption that the multi-angle instability plays no role and that 
the neutrino ensemble is largely characterized by self-maintained 
coherence. In this case one is led to a unique formulation of the 
collective equations of motion that imply that collective flavor 
oscillations should be thought of as a one-dimensional phenomenon 
along the streamlines of the underlying neutrino flux. Close to the 
source these streamlines are usually curved even though, of course, 
the underlying neutrino trajectories are straight. (We have neglected 
the gravitational bending of trajectories.) Therefore, even if the 
neutrino stream has no global symmetries, the collective oscillation 
problem is relatively simple. 
 
We have used the concept of ``self-maintained coherence'' in the most 
restrictive sense that applies when the neutrino gas is dense, i.e., 
when a typical neutrino-neutrino interaction energy $\mu$ is large 
compared to a typical vacuum oscillation frequency 
$\omega_0=\langle\Delta m^2/2E\rangle$. The neutrino ensemble in this 
case evolves along a streamline as one unit that previously has been 
described as a gyroscopic pendulum in flavor space. All neutrino and 
anti-neutrino polarization vectors point essentially in the same 
direction in flavor space, the pendulum direction, allowing for the 
simplifications that lead to our collective equations. We have 
provided a prescription for defining the effective neutrino-neutrino 
interaction strength $\mu$ that works for general source geometries. 
Our result is somewhat different from what was used in the 
previous literature, but our expression for $\mu$
agrees better with numerical simulations. 
 
There are more general forms of collective behavior. In 
a homogeneous isotropic system with a density that decreases 
adiabatically in time, the polarization vectors are at first aligned 
and stay in a single co-rotating plane even when $\mu$ becomes of 
order $\omega_0$ and smaller. However, they will align or anti-align 
themselves in the mass direction, leading to the phenomenon of 
spectral splits. This is a more general case of self-maintained 
coherence beyond our simple picture and is analytically fully 
understood. 
In the spherically symmetric case, numerical multi-angle
simulations show a similar behavior.
If the asymmetry between neutrinos and antineutrinos is large enough to 
prevent multi-angle decoherence, the neutrino and antineutrino modes 
evolve collectively essentially along the co-rotating plane, but with 
intriguing three-dimensional patterns that evolve collectively and 
are stable in flavor space. Once more spectral splits develop, 
similar to the single-angle approximation. 
 
We speculate that these more general forms of behavior also occur for
more general source geometries if we follow the neutrino stream
lines. It is clear that multi-angle decoherence will occur if the
asymmetry between neutrinos and anti-neutrinos is extremely small.
Even a few per cent asymmetry, on the other hand, implies that close
to the source we have synchronized oscillations, or, in the pendulum
picture, it precesses fast without nutations. Therefore, close to the
source where the streamlines are curved, our treatment should be most
appropriate. If the effective mixing angle is small (as it will be
because of the presence of ordinary matter), superficially no flavor
oscillations happen close to the source and the ``true action'' begins
at the synchronization radius where the inverted pendulum begins to
nutate and begins to lose its initial orientation.  If this point is
reached at some distance, the streamlines will essentially point
straight away from the source and naively one should think that
henceforth the evolution is similar to the spherically symmetric
case. In fact in the adiabatic approximation, one only needs to know
the local value of $\omega_0/\mu$ in order to determine the neutrino
survival probabilities, for any source geometry.
 
Granting these assumptions, we have briefly studied a toy model for
coalescing neutron stars where the source is taken to be a
homogeneously radiating disk. For our numerical example, we find that
the region where collective oscillations will convert the original
$\nu_e$ and ${\bar\nu_e}$ fluxes is quite far from the emission
surface. Therefore these flavor conversions should have a negligible
effect on the neutrino energy deposition in the plasma. However, due
to the uncertainties of the original neutrino fluxes, we can not
exclude situations with a smaller asymmetry between the original
fluxes. Such tiny neutrino asymmetries may give rise to multi-angle
decoherence, that could produce significant flavor conversions near
the emitting disk.  This question ultimately will need to be addressed
numerically by means of large-scale numerical simulations of neutrino
flavor evolution near a non-spherical source, in analogy to what has
been done in the case of neutrinos streaming off a SN core.
 
Our study suggests that self-maintained coherence among different
neutrino modes may well be a typical form of behavior even in
non-spherical systems and notably in the interesting case of
coalescing neutron stars. The final verdict on the role of collective
neutrino oscillations in such systems can only come from numerical
studies. At very least, our approach provides a simple limiting case
against which one can measure the output of numerical simulations.
 
 
\begin{acknowledgments}
A.D.\ and B.D.\ would like to thank K.~Damle, R.~Loganayagam, and
S.~Raychaudhuri for useful discussions. In Munich, this work was
partly supported by the Deutsche Forschungsgemeinschaft (grant TR-27
``Neutrinos and Beyond''), by the Cluster of Excellence ``Origin and
Structure of the Universe'' and by the European Union (contract
No.\ RII3-CT-2004-506222). In Mumbai, partial support by a Max Planck
India Partnergroup grant is acknowledged. A.M.\ acknowledges support
by the Italian Istituto Nazionale di Fisica Nucleare (INFN) through a
post-doctoral fellowship.
\end{acknowledgments} 
 
\appendix 
\section{Disk source}                 \label{app:stream} 

We here briefly describe the numerical integrations for a disk source
that were used in the main text.  The direction of neutrino momenta
are determined by two polar angles $\theta_{p}$ and $\phi_p$, by which
we can express the component of the neutrino velocity in $x$ and $z$
direction respectively~by
\begin{eqnarray} 
v^x &=& \cos\phi_p\sin\theta_p  \nonumber  \,,\\ 
v^z &=& \cos\theta_p \,. 
\end{eqnarray} 
 
Our first task is to determine the flux ${\bf F}_{\bf x}$, i.e., the
$x$ and $z$ components $F^x_{x,z}$ and $F^z_{x,z}$, each as a function
of $x$ and $z$.  These are given by
\begin{eqnarray} 
F^x_{x,z} &=& \frac{1}{8 \pi^3} 
\int \D E\, \D\cos\theta_p\,\D\phi_p v^x f_{x,z}(v^x,v^z) \nonumber \,,\\ 
F^z_{x,z} &=& \frac{1}{8 \pi^3} 
\int \D E\, \D\cos\theta_p\,\D\phi_p v^z f_{x,z}(v^x,v^z)  \,.
\end{eqnarray} 
 
One can also define the local neutrino density 
\begin{equation}
n_{x,z} = \frac{1}{8 \pi^3} 
\int \D E \D\cos\theta_p \D\phi_p  f_{x,z}(v^x,v^z)  \,\ .  
\end{equation}
 
In the case of synchronization, from Eq.~(\ref{eq:EOM4g}) 
we arrive at the following EOM 
\begin{equation}
(\langle v^x_{x,z} \rangle \partial_x + 
\langle v^z_{x,z} \rangle \partial_z) a_{x,z} = \omega_0 \,\ , 
\end{equation}
where 
\begin{eqnarray} 
\langle v^x_{x,z} \rangle &=& \frac{F^x_{x,z}}{n_{x,z}} \,\ , \nonumber\\ 
\langle v^z_{x,z} \rangle  &=& \frac{F^z_{x,z}}{n_{x,z}} \,\ . 
\end{eqnarray} 
In this case, we have obtained a first order partial differential equation 
(PDE). 
Since  $\langle v^x_{x,z} \rangle$ and  $\langle v^z_{x,z} \rangle$ are not constant, 
in this case the streamlines  will be curved. 
If the streamlines are graphs of the function $z(x)$, it follows 
\begin{equation}
\frac{\D z}{\D x} = \frac{\langle v^z_{x,z}\rangle} 
{\langle v^x_{x,z} \rangle} \,\ , 
\label{eq:ode} 
\end{equation}
(i.e., the tangent line to the graph of $z(x)$ at $(x,z)$ is parallel 
to ${\bf v}$.) 
The ordinary differential equation (ODE) in Eq.~(\ref{eq:ode}) 
is the so-called characteristic equation for the associated PDE. 
Its solutions are the streamlines  for the PDE. 
 
\begin{figure} 
\begin{center} 
\epsfig{figure=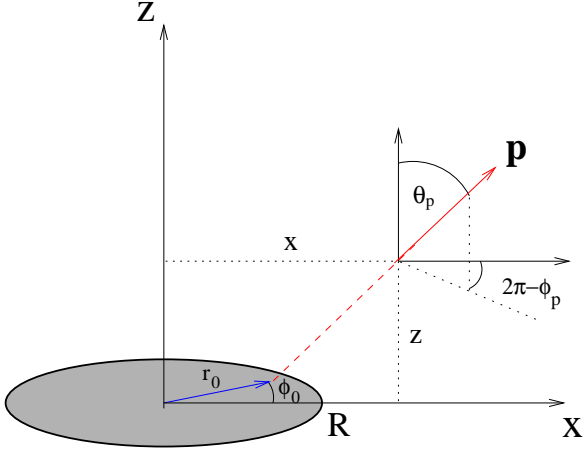,width=0.9\columnwidth,angle=0} 
\end{center} 
\caption{Relation between the momenta variables $\theta_p$ and $\phi_p$ and  
 coordinates $r_0$ and $\phi_0$ on the disk.
\label{fig:thetapphip}} 
\end{figure} 
 
 The range of the previous angular integrations are determined 
by the coordinates $x$ and $z$. It could 
be convenient to express the previous integrals at a given 
 point, writing $\theta_p$ and $\phi_p$ in terms of the 
 coordinates $r_0$ and $\phi_0$ on the disk, by tracing back the 
direction of neutrino momenta on the disk, as shown in 
Fig.~\ref{fig:thetapphip}. 
Let the neutrino emitted from a point 
$(r_0 \cos \phi_0, r_0 \sin \phi_0, 0)$ 
pass through the location $(x,0,z)$.
The direction of the momentum of this neutrino is along the
unit vector
$(\sin \theta_p \cos \phi_p, \sin \theta_p \sin \phi_p, \cos \theta_p)$.
Therefore,
\begin{eqnarray}
(x-r_0 \cos \phi_0, -r_0 \sin \phi_0, z) \propto  
\phantom{morespace}  & &\nonumber \\
\phantom{space} 
(\sin \theta_p \cos \phi_p, \sin \theta_p \sin \phi_p, \cos \theta_p) \; .& &
\end{eqnarray}
We can now write down $\theta_p$ and $\phi_p$ explicitly
in terms of $r_0, \phi_0, x, z$:
\begin{equation}
\tan \theta_p = \frac{\sqrt{r_0^2 + x^2 -2 x r_0 \cos \phi_0}}{z} \; ,
\end{equation}
so that 
\begin{eqnarray}
\cos \theta_p &=& \frac{z}{\sqrt{r_0^2 + x^2 - 2 x r_0 \cos \phi_0 + z^2}}   \,\ , \nonumber \\
\sin\theta_p &=& \frac{\sqrt{r_0^2 +x^2 - 2 xr_0 \cos \phi_0}} 
{\sqrt{r_0^2 +x^2 -2 xr_0 \cos\phi_0 + z^2}}  \; .
\end{eqnarray}

For the angle $\phi_p$, we get
\begin{eqnarray} 
\sin \phi_p &=& -\frac{r_0 \sin \phi_0}{\sin \theta_p} 
= -\frac{r_0 \sin \phi_0}{z \tan \theta_p} \,\ , \nonumber\\ 
\cos \phi_p &=& \frac{x-r_0\cos\phi_0}{\sin\theta_p} 
= \frac{x-r_0\cos\phi_0}{z \tan \theta_p} \,\ ,
\end{eqnarray} 
so that 
\begin{equation}
 \phi_p = - \arctan\left(\frac{r_0 \sin\phi_0}{x-r_0 \cos \phi_0}\right) \,\ . 
\end{equation}
Using as variables of integration $r_0$ and $\phi_0$, 
we obtain 
\begin{eqnarray} 
n_{x,z}&=& 
\int \frac{\D E}{8 \pi^3} \D \cos\theta_p \D \phi_p f_{x,z}(v^x,v^z) \nonumber \\ 
&=& 
\int \D r_0 \D \phi_0 \textrm{det}J(r_0,\phi_0) {\tilde f}_{x,z}(v^x,v^z)  \,\ , 
\end{eqnarray} 
where $J(r_0,\phi_0)$ is the Jacobian that relates the two coordinate systems, and 
$0 \leq r_0\leq 1$ and $0\leq \phi_0 \leq 2 \pi$. 
 
The derivatives relevant for the Jacobian are the following 
\begin{eqnarray} 
\frac{\partial \cos\theta_p}{\partial r_0} &=& \frac{-z(r_0-x\cos\phi_0)} 
{(r_0^2 + x^2 - 2 xr_0 \cos\phi_0 + z^2)^{3/2}} \,\ , \nonumber\\ 
\frac{\partial \cos \theta_p}{\partial \phi_0} &=& \frac{-zxr_0\sin\phi_0} 
{(r_0^2 +x^2 - 2 xr_0\cos\phi_0 
+z^2)^{3/2}} \,\ , \nonumber  \\ 
\frac{\partial \phi_p}{\partial r_0}&=&-\frac{x \sin \phi_0}{r_0^2 + x^2 -2xr_0\cos\phi_0} \,\ ,
 \nonumber \\ 
\frac{\partial \phi_p}{\partial \phi_0}&=& \frac{r_0^2 -xr_0\cos\phi_0}{r_0^2 +x^2 -2xr_0\cos\phi_0} \,\ , 
\end{eqnarray} 
so that 
\begin{eqnarray} 
\textrm{det}J &=& 
\frac{\partial \cos\theta_p}{\partial r_0} \frac{\partial \phi_p}{\partial \phi_0} 
- \frac{\partial \cos \theta_p}{\partial \phi_0} \frac{\partial \phi_p}{\partial r_0} 
 \nonumber \\ 
&=& 
\frac{-zr_0}{(r_0^2 +x^2 - 2 xr_0\cos\phi_0 
+z^2)^{3/2}}  \; .
\end{eqnarray} 
The net result is
\begin{eqnarray} 
F^x_{x,z} &=& 
\int \frac{\D E}{8 \pi^3}  \D \phi_0 \D r_0 \frac{-zr_0(x-r_0 \cos\phi_0) f_{x,z}(v^x,v^z)} 
{(r_0^2 +x^2-2xr_0\cos\phi_0 +z^2)^2} \,\ , \nonumber \\ 
F^z_{x,z} &=& 
\int \frac{\D E}{8 \pi^3}  \D \phi_0 \D r_0 \frac{z(-zr_0) 
f_{x,z}(v^x,v^z)}{(r_0^2 +x^2-2xr_0\cos\phi_0 +z^2)^2} 
 \,\ .\nonumber \\
\end{eqnarray}



\begin{thebibliography}{00} 
 
\bibitem{Pantaleone:1992eq} 
  J.~Pantaleone, 
  ``Neutrino oscillations at high densities,'' 
  Phys.\ Lett.\ B {\bf 287}, 128 (1992). 
 
\bibitem{Sigl:1992fn} 
  G.~Sigl and G.~Raffelt, 
  ``General kinetic description of relativistic mixed neutrinos,'' 
  Nucl.\ Phys.\ B {\bf 406}, 423 (1993). 
 
\bibitem{Samuel:1993uw} 
  S.~Samuel, 
  ``Neutrino oscillations in dense neutrino gas\-es,'' 
  Phys.\ Rev.\ D {\bf 48}, 1462 (1993). 
 
\bibitem{Kostelecky:1993dm} 
  V.~A.~Kosteleck\'y and S.~Samuel, 
  ``Neutrino oscillations in the early universe with an inverted 
  neutrino mass hierarchy,'' 
  Phys.\ Lett.\ B {\bf 318}, 127 (1993). 
 
\bibitem{Kostelecky:1995dt} 
  V.~A.~Kosteleck\'y and S.~Samuel, 
  ``Self-maintained coherent oscillations in dense neutrino gases,'' 
  Phys.\ Rev.\ D {\bf 52}, 621 (1995) 
  [hep-ph/9506262]. 
 
\bibitem{Samuel:1996ri} 
  S.~Samuel, 
  ``Bimodal coherence in dense selfinteracting neutrino gases,'' 
  Phys.\ Rev.\ D {\bf 53}, 5382 (1996) 
  [hep-ph/9604341]. 
 
\bibitem{Pastor:2001iu} 
  S.~Pastor, G.~G.~Raffelt and D.~V.~Semikoz, 
  ``Physics of synchronized neutrino oscillations caused by 
  self-interactions,'' 
  Phys.\ Rev.\ D {\bf 65}, 053011 (2002) 
  [hep-ph/0109035]. 
 
\bibitem{Wong:2002fa} 
  Y.~Y.~Y.~Wong, 
  ``Analytical treatment of neutrino asymmetry equilibration 
  from flavour oscillations in the early universe,'' 
  Phys.\ Rev.\ D {\bf 66}, 025015 (2002) 
  [hep-ph/ 0203180]. 
 
\bibitem{Abazajian:2002qx} 
  K.~N.~Abazajian, J.~F.~Beacom and N.~F.~Bell, 
  ``Stringent constraints on cosmological neutrino 
   antineutrino asymmetries 
   from synchronized flavor transformation,'' 
  Phys.\ Rev.\  D {\bf 66}, 013008 (2002) 
  [astro-ph/0203442]. 
 
\bibitem{Pastor:2002we} 
  S.~Pastor and G.~Raffelt, 
  ``Flavor oscillations in the supernova hot bubble region: 
  Nonlinear  effects of neutrino background,'' 
  Phys.\ Rev.\ Lett.\  {\bf 89}, 191101 (2002) 
  [astro-ph/0207281]. 
 
\bibitem{Sawyer:2004ai} 
  R.~F.~Sawyer, 
  ``Classical instabilities and quantum speed-up in the evolution of 
  neutrino clouds,'' 
  hep-ph/0408265. 
 
\bibitem{Sawyer:2005jk} 
  R.~F.~Sawyer, 
  ``Speed-up of neutrino transformations in a supernova environment,'' 
  Phys.\ Rev.\  D {\bf 72}, 045003 (2005) 
  [hep-ph/0503013]. 
 
\bibitem{Sawyer:2008zs} 
  R.~F.~Sawyer, 
  ``The multi-angle instability in dense neutrino systems,'' 
  arXiv:0803.4319 [astro-ph]. 
 
 
\bibitem{Duan:2005cp} 
  H.~Duan, G.~M.~Fuller and Y.~Z.~Qian, 
  ``Collective neutrino flavor transformation in supernovae,'' 
  Phys.\ Rev.\  D {\bf 74}, 123004 (2006) 
  [astro-ph/0511275]. 
 
\bibitem{Duan:2006an}
    H.~Duan, G.~M.~Fuller, J.~Carlson and Y.~Z.~Qian, 
    ``Simulation of coherent non-linear neutrino flavor 
    transformation in the supernova environment. I: Correlated 
    neutrino trajectories,'' Phys.\ Rev.\ D {\bf 74}, 105014 (2006) 
    [astro-ph/0606616]. 
 
\bibitem{Hannestad:2006nj} 
  S.~Hannestad, G.~G.~Raffelt, G.~Sigl and Y.~Y.~Y.~Wong, 
  ``Self-induced conversion in dense neutrino gases: 
  Pendulum in flavor space,'' 
  Phys.\ Rev.\ D {\bf 74}, 105010 (2006); 
  Erratum ibid.\ {\bf 76}, 029901 (2007) 
  [astro-ph/0608695]. 
 
\bibitem{Raffelt:2007yz} 
  G.~G.~Raffelt and G.~Sigl, 
  ``Self-induced decoherence in dense neutrino gases,'' 
  Phys.\ Rev.\ D {\bf 75}, 083002 (2007) 
  [hep-ph/0701182]. 
 
\bibitem{Duan:2007mv} 
  H.~Duan, G.~M.~Fuller, J.~Carlson and Y.~Z.~Qian, 
  ``Analysis of collective neutrino flavor transformation in 
  supernovae,'' 
  Phys.\ Rev.\  D {\bf 75}, 125005 (2007) 
  [astro-ph/0703776]. 
 
\bibitem{Raffelt:2007cb} 
  G.~G.~Raffelt and A.~Yu.~Smirnov, 
  ``Self-induced spectral splits in supernova neutrino fluxes,'' 
  Phys.\ Rev.\  D {\bf 76}, 081301 (2007); 
  Erratum ibid.\ {\bf 77}, 029903 (2008) 
  [arXiv:0705.1830 (hep-ph)]. 
 
\bibitem{Raffelt:2007xt} 
 G.~G.~Raffelt and A.~Yu.~Smirnov, 
 ``Adiabaticity and spectral splits in collective neutrino 
 transformations,'' 
 Phys.\ Rev.\ D {\bf 76}, 125008 (2007) 
 [arXiv:0709.4641 (hep-ph)]. 
 
\bibitem{EstebanPretel:2007ec} 
 A.~Esteban-Pretel, S.~Pastor, R.~Tom\`as, 
 G.~G.~Raffelt and G.~Sigl, 
 ``Decoherence in supernova neutrino transformations 
 suppressed by deleptonization,'' 
 Phys.\ Rev.\ D {\bf 76}, 125018 (2007) 
[arXiv:0706.2498 (astro-ph)]. 
 
\bibitem{Duan:2007fw} 
 H.~Duan, G.~M.~Fuller and Y.~Z.~Qian, 
 ``A simple picture for neutrino flavor transformation 
 in supernovae,'' 
 Phys.\ Rev.\  D {\bf 76}, 085013 (2007) 
 [arXiv:0706.4293 (astro-ph)]. 
 
\bibitem{Duan:2007bt} 
 H.~Duan, G.~M.~Fuller, J.~Carlson and Y.Z.~Qian, 
 ``Neutrino mass hierarchy and stepwise spectral swapping of supernova 
 neutrino flavors,'' 
 Phys.\ Rev.\ Lett. {\bf 99}, 241802 (2007) 
 [arXiv:0707.0290 (astro-ph)]. 
 
\bibitem{Fogli:2007bk} 
 G.~L.~Fogli, E.~Lisi, A.~Marrone and A.~Mirizzi, 
 ``Collective neutrino flavor transitions in supernovae and the role of 
 trajectory averaging,'' 
 JCAP {\bf 0712}, 010 (2007) 
 [arXiv:0707.1998 (hep-ph)]. 
 
\bibitem{Duan:2007sh}
  H.~Duan, G.~M.~Fuller, J.~Carlson and Y.~Z.~Qian,
  ``Flavor evolution of the neutronization neutrino burst from an O-Ne-Mg
  core-collapse supernova,''
  Phys.\ Rev.\ Lett.\  {\bf 100}, 021101 (2008)
  [arXiv:0710.1271 (astro-ph)].
 
\bibitem{Dasgupta:2008cd} 
  B.~Dasgupta, A.~Dighe, A.~Mirizzi and G.~G.~Raffelt, 
  ``Spectral split in prompt supernova neutrino burst: 
  Analytic three-flavor treatment,'' 
  arXiv:0801.1660 [hep-ph]. 
 
\bibitem{EstebanPretel:2007yq} 
  A.~Esteban-Pretel, S.~Pastor, R.~Tom\`as, G.~G.~Raffelt and G.~Sigl, 
  ``Mu-tau neutrino refraction and collective three-flavor 
  transformations in supernovae,'' 
  Phys.\ Rev.\  D {\bf 77}, 065024 (2008) 
  [arXiv:0712.1137 (astro-ph)]. 
 
\bibitem{Dasgupta07} 
 B.~Dasgupta and A.~Dighe 
 ``Collective three-flavor oscillations of supernova neutrinos,'' 
 [arXiv:0712.3798 (hep-ph)]. 
 
\bibitem{Duan:2008za}
  H.~Duan, G.~M.~Fuller and Y.~Z.~Qian,
  ``Stepwise spectral swapping with three neutrino flavors,''
  Phys.\ Rev.\  D {\bf 77}, 085016 (2008)
  [arXiv:0801.1363 (hep-ph)].
 
\bibitem{Dasgupta:2008my} 
  B.~Dasgupta, A.~Dighe and A.~Mirizzi, 
  ``Identifying neutrino mass hierarchy at extremely small 
  theta(13) through Earth matter effects in a supernova signal,'' 
  arXiv:0802.1481 [hep-ph]. 
 
\bibitem{Duan:2008eb} 
  H.~Duan, G.~M.~Fuller and J.~Carlson, 
  ``Simulating nonlinear neutrino flavor evolution,'' 
  arXiv:0803.3650 [astro-ph]. 
 
\bibitem{Ruffert:1998qg} 
  M.~Ruffert and H.-T.~Janka, 
  ``Gamma-ray bursts from accreting black holes in 
  neutron star mergers,'' 
  Astron.\ Astrophys.\ {\bf 344}, 573 (1999) 
  [astro-ph/9809280]. 
 
 
\bibitem{Dolgov:1980cq} 
  A.~D.~Dolgov, 
  ``Neutrinos in the early universe,'' 
  Yad.\ Fiz.\  {\bf 33}, 1309 (1981) 
  [Sov.\ J.\ Nucl.\ Phys.\  {\bf 33}, 700 (1981)]. 
 
\bibitem{Rudzsky:1990} 
  M.~A.~Rudzsky, 
  ``Kinetic equations for neutrino spin- and type-oscillations 
  in a medium,'' 
  Astrophys. Space Science {\bf 165}, 65 (1990). 
  
\bibitem{mckellar&thomson} 
  B.~H.~J.~McKellar and M.~J.~Thomson, 
  ``Oscillating doublet neutrinos in the early universe,'' 
  Phys.\ Rev.\ D {\bf 49}, 2710 (1994). 
 
\bibitem{Strack:2005ux}
  P.~Strack and A.~Burrows,
  ``A generalized Boltzmann formalism for oscillating neutrinos,''
  Phys.\ Rev.\  D {\bf 71}, 093004 (2005)
  [hep-ph/0504035].
 
\bibitem{Cardall:2007zw} 
  C.~Y.~Cardall, 
  ``Liouville equations for neutrino distribution matrices,'' 
  arXiv:0712.1188 [astro-ph]. 
 
\bibitem{Mrowczynski:2007hb} 
  S.~Mrowczynski and M.~H.~Thoma, 
  ``What do electromagnetic plasmas tell us about quark-gluon 
  plasma?,'' 
  Ann.\ Rev.\ Nucl.\ Part.\ Sci.\  {\bf 57}, 61 (2007) 
  [nucl-th/0701002]. 
 
\bibitem{pde} 
E.~C.~Zachmanoglou and D.~W.~Thoe, 
{\em Introduction to Partial Differential 
Equations with Applications}
(Dover Publications, New York, 1986). 
 
\bibitem{goodman} 
J.~Goodman, A.~Dar, and S.~Nussinov, 
``Neutrino annihilation in Type II supernovae,'' 
Astrophys.\ J.\ Lett. {\bf 314}, L7 (1987). 
 
\bibitem{Volkas:1999gb} 
  R.~R.~Volkas and Y.~Y.~Y.~Wong, 
  ``Matter-affected neutrino oscillations in ordinary and mirror stars and 
  their implications for gamma-ray bursts,'' 
  Astropart.\ Phys.\  {\bf 13}, 21 (2000) 
  [astro-ph/9907161]. 
 
\end{thebibliography}
\end{document}